\documentclass[sigconf,nonacm]{acmart}
\AtBeginDocument{%
  }

\usepackage{xcolor}
\usepackage{tcolorbox}
\usepackage{tabularx}
\usepackage{balance}
\usepackage{enumitem}
\setcopyright{acmlicensed}
\copyrightyear{2018}
\acmYear{2018}
\acmDOI{XXXXXXX.XXXXXXX}
\acmConference[Conference acronym 'XX]{Make sure to enter the correct
  conference title from your rights confirmation email}{June 03--05,
  2018}{Woodstock, NY}
\acmISBN{978-1-4503-XXXX-X/2018/06}

\begin{document}

\title{Forecasting Clicks in Digital Advertising: Multimodal Inputs and Interpretable Outputs}

\author{Briti Gangopadhyay}
\email{briti.gangopadhyay@sony.com}
\affiliation{%
  \institution{Sony Group Corporation}
  % \city{Hekla}
  \country{Japan}
  }

\author{Zhao Wang}
\email{zhao.wang@sony.com}
\affiliation{%
  \institution{Sony Group Corporation}
  \country{Japan}
}

\author{Shingo Takamatsu}
\email{shingo.takamatsu@sony.com}
\affiliation{%
  \institution{Sony Group Corporation}
  \country{Japan}
}

\renewcommand{\shortauthors}{Trovato et al.}

\begin{abstract}
Forecasting click volume is a key task in digital advertising, influencing both revenue and campaign strategy. Traditional time series models rely solely on numerical data, often overlooking rich contextual information embedded in textual elements, such as keyword updates. We present a multimodal forecasting framework that combines click data with textual logs from real-world ad campaigns and generates human-interpretable explanations alongside numeric predictions. Reinforcement learning is used to improve comprehension of textual information and enhance fusion of modalities. Experiments on a large-scale industry dataset show that our method outperforms baselines in both accuracy and reasoning quality.
\end{abstract}

  %% \includegraphics[width=\textwidth
  %% {sampleteaser}
%%  \caption{Seattle Mariners at Spring %%Training, 2010.}
 %% \Description{Enjoying the baseball game %%from the third-base
 %% seats. Ichiro Suzuki preparing to bat.}
 %% \label{fig:teaser}

%%\received{20 Feruary 2007}
%%\received[revised]{12 March 2009}
%%\received[accepted]{5 June 2009}

\maketitle

\section{Introduction}
\begin{figure}[h]
  \centering
  \includegraphics[width=\linewidth]{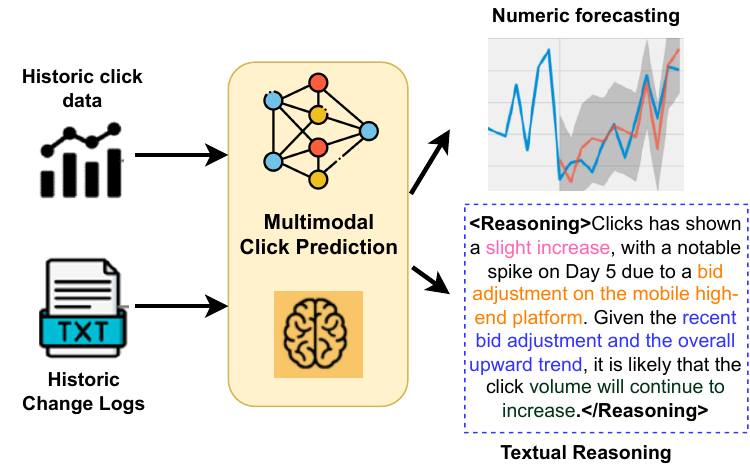}
  \caption{Multimodal click prediction leveraging historical click data and text-based change logs. The forecaster provides both the prediction and a text-based explanation.}
  \label{teaser}
\end{figure}
In the realm of digital advertising, gauging and adapting to click trends is essential for optimizing marketing strategies \cite{forbes2025clicks, wang2017deep, greenlaneClickCurve, guo2017deepfm, zhou2018deep}. This task typically falls under the domain of time series forecasting (TSF). Traditional TSF approaches often rely on unimodal data \cite{zhou2021informer, liu2023itransformer, zhou2022film, wang2017deep}, thereby missing the opportunity to extract semantic information embedded in text-based events such as ad copy updates or keyword additions and removals.

Multimodal Time Series Forecasting (MMTSF) has emerged as a promising research direction, propelled by the recent advances in large language models (LLMs). Studies such as \cite{gruver2023large, liu2024lstprompt} explore the zero-shot forecasting capabilities of LLMs, while also highlighting performance limitations on complex datasets and dependencies on specific model versions, which hinder practical deployment. MMTSF is particularly relevant in digital advertising and recommendation systems \cite{yuan2023go}. For instance, \cite{li2024multi} employed MMTSF for forecasting new product sales, while \cite{huan2024exploring} introduced a large-scale multimodal dataset for click-through rate (CTR) prediction in recommendation systems. Recently, \cite{liu2024time} proposed a large-scale dataset and the first general-purpose MMTSF architecture, integrating both textual and numerical data across diverse domains. Building upon this line of work, our paper introduces a MMTSF with Reinforcement Learning (RL)-based fine-tuning tailored to the digital advertising domain. We demonstrate that incorporating RL not only improves forecasting performance, but also generates meaningful textual cues enabling advertisers to adapt the evolving and dynamic nature of ad ecosystems. In this paper, we make the following key contributions:

\begin{itemize}
    \item A custom reward function to fine-tune LLM for improved text understanding in click forecasting context.
    
    \item An end-to-end pipeline for multimodal click forecasting that not only predicts numerical click trends but also generates textual explanations for its predictions as illutrated in Fig \ref{teaser}. 
    \item Empirical studies that demonstrate the effectiveness of our approach, highlighting improvements over traditional and multimodal baselines.
\end{itemize}

\begin{figure*}[t]
    \centering
    \begin{tcolorbox}[colback=yellow!5!white,colframe=yellow!75!black,title=Prompt Template For LLM,fonttitle=\bfseries]
    \small
    \textbf{Prompt:}

    \vspace{4pt}
    \textit{You are an expert in data analysis and forecasting. I will provide you with a time series of rolling averages of daily clicks for a campaign, recent change logs, the type of ad being delivered, and the bidding strategy.}

    \vspace{4pt}
    \textbf{Inputs:} 1. \textbf{Rolling average of clicks (past \emph{l} days)}: \{\textit{Numeric data}\} 
    
    2. \textbf{Change logs (past \emph{l} days)}: \{\textit{Textual data}\} 
    
    3. \textbf{Ad type}: \{\textit{Ad category}\} 
    
    4. \textbf{Bidding strategy}: \{\textit{Strategy type}\}
    
    \vspace{4pt}
    \textbf{Task:} Analyze the data and provide a **concise** two-sentence reasoning. Provide **exactly one-word** as the prediction (Increase/Decrease) for click trend on average over next \emph{\{h\}} days. Format your response strictly as follows:
    
    {\color{blue}\texttt{<Reasoning> Your reasoning sentence </Reasoning>}
    \texttt{<Prediction> Increase/Decrease </Prediction>}}
    
    \end{tcolorbox}
    \caption{Prompt utilised for RL-based LLM finetuning.}
    \label{fig::multimodal_prompt}
\end{figure*}

\section{Task Description}

A general TSF task involves predicting future event trends based on historical data. Formally, let the numerical input be denoted by $\mathcal{X} \in \mathbb{R}^{l \times d_{\text{num}}}$, where $l$ is the \emph{lookback window}, determined by domain-specific requirements, and $d_{\text{num}}$ is the number of numerical features. The forecasting model aims to generate an output over a future horizon of length $h$, denoted as $\mathcal{Y} \in \mathbb{R}^{h \times d_{\text{out}}}$.

In addition to the numerical features, the input includes textual features represented as $\mathcal{C} \in \mathbb{R}^{l \times d_{\text{txt}}}$. The multimodal forecasting model, parameterized by $\theta$, is denoted as $M_{\theta} : \mathcal{X} \times \mathcal{C} \rightarrow \mathcal{Y}$.

At each time step $t$, a training sample is defined as $(X_t + C_t, Y_t)$, where:
$
(X_t, C_t) = \left[
(x_{t-l}, c_{t-l}), 
(x_{t-l+1}, c_{t-l+1}), \dots, 
(x_{t}, c_{t})
\right]
$,
$
Y_t = [y_{t+1}, y_{t+2}, \dots, y_{t+h}]
$. Additionally, an LLM, denoted $\mathcal{L}_{\omega}$, is used to generate a supportive textual counterpart $y_{\text{text}_{t+i}}$.

\begin{figure}[h]
  \centering
  \includegraphics[width=\linewidth]{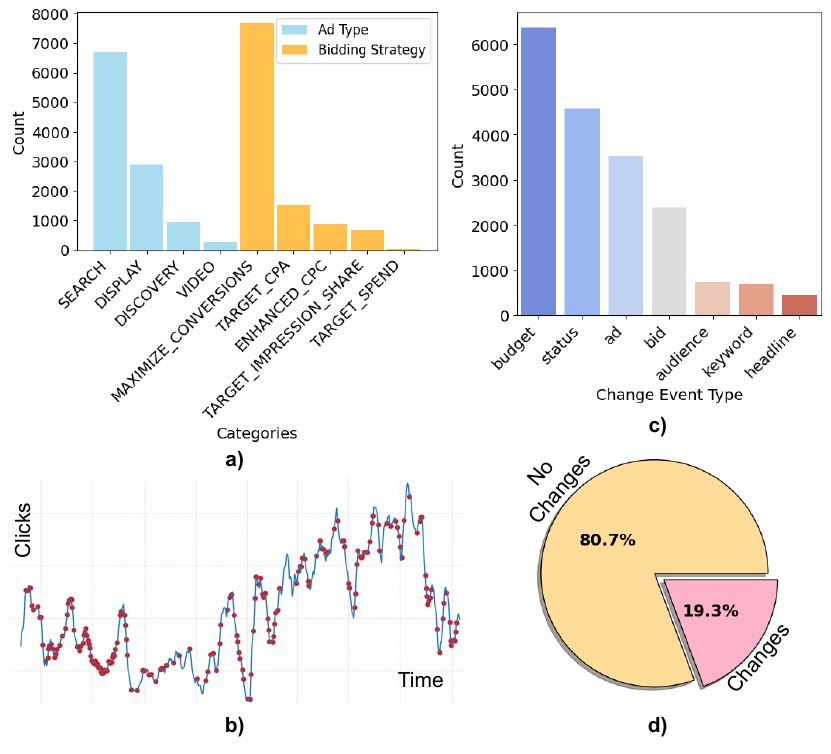}
  \caption{
a) Distribution of campaign types and bidding strategies (CPA = Cost per Acquisition, CPC = Cost per Click). 
b) Click trend of a representative campaign; red dots denote text change events. 
c) Distribution of various text-based events. 
d) Change versus no-change instances in the dataset.
}
  \label{datadescribe}
\end{figure}

\subsection{Dataset Description}

We obtain data from 46 real-world advertisement campaigns, each containing both numerical time series data and corresponding change logs. These change logs consist of textual information describing various configuration changes within the campaigns. Examples of such changes include budget adjustments, keyword additions and deletions, ad headline modifications, bid strategy changes, and bid value updates. The frequency distribution of different change events in the dataset is illustrated in Fig.~\ref{datadescribe}c. The campaigns span multiple products and a variety of campaign types, including Search, Display, Discovery, and Video. To construct a multimodal dataset, we align the textual change logs with the numerical time series data based on the event timestamps. Dates on which no events occurred are explicitly marked as "no changes". As shown in Fig.~\ref{datadescribe}d, the textual data is extremely sparse, making it a weak signal when directly integrated with a TSF model. To address this, we use LLM summaries that can better leverage such sparse textual signals. In total, 44 campaigns comprising 10,798 data points are used for training, while two held-out campaigns with 1,045 previously unseen data points are reserved for testing.

\begin{figure*}[h]
  \centering
  \includegraphics[width=0.8\linewidth]{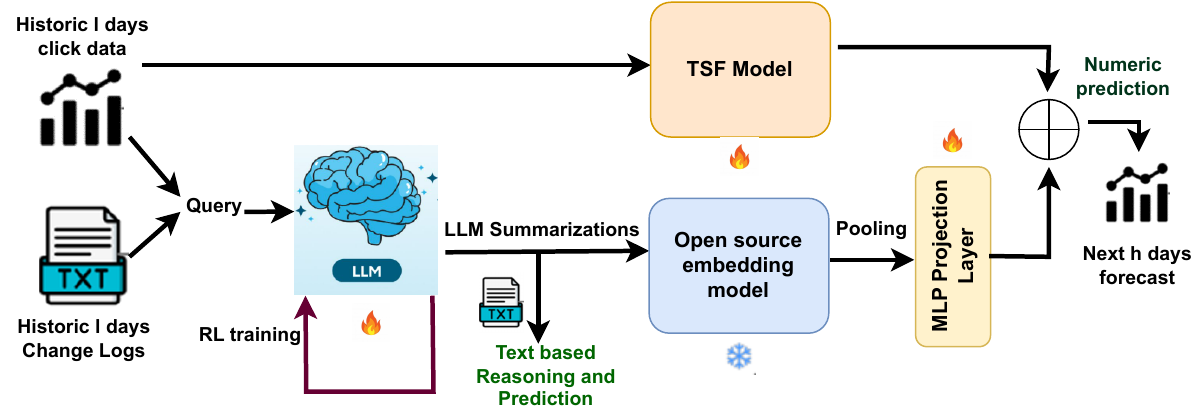}
  \caption{Multimodal click forecasting pipeline. LLM Summerizations are <Reasoning> + <Prediction> output from the LLM.}
  \label{modelarchi}
\end{figure*}

\subsection{RL-based Fine-tuning}

The training data is converted into prompts for fine-tuning the LLM, as illustrated in Fig.~\ref{fig::multimodal_prompt}. We design the following reward function to encourage accurate and well-structured summarization:

\begin{equation}
    R = S_{\text{format}} + \mathbb{I}(\hat{y} = y) + \mathbb{I}(s(r) = y) \cdot c
\end{equation}

\noindent where:
\begin{itemize}[leftmargin=*, nosep]
    \item \( R \) is the total reward.
    \item \( S_{\text{format}} \) is the format compliance score, penalized by up to \(-0.5\) for missing or incorrect \texttt{<prediction>} or \texttt{<reasoning>} tags.
    \item \( \mathbb{I}(\hat{y} = y) \) is the prediction accuracy term, defined as:
    \[
    \mathbb{I}(\hat{y} = y) = 
    \begin{cases}
        1, & \text{if the predicted value matches the ground truth,} \\[4pt]
        0, & \text{otherwise.}
    \end{cases}
    \]
    \item \( \mathbb{I}(s(r) = y) \cdot c \), where \( s(r) \) denotes the sentiment inferred from the generated reasoning. This term evaluates whether the sentiment aligns with the actual trend and scales it by the sentiment model’s confidence score \( c \). The sentiment alignment condition is:
    \[
    (\text{sentiment} = \text{POSITIVE} \land \text{actual} = \text{increase}) \;\lor
    \]
    \[
    (\text{sentiment} = \text{NEGATIVE} \land \text{actual} = \text{decrease})
    \]
\end{itemize}

The reward function thus evaluates the model's output based on three key components: format compliance, prediction accuracy, and reasoning alignment. Without sentiment alignment, we find that the model's reasoning often contradicts the prediction. We adopt Group Relative Policy Optimization (GRPO) \cite{guo2025deepseek, shao2024deepseekmath} as the training method, which avoids the need for a separately trained critic, making it well suited for fine-tuning under limited computational resources. Sentiment and confidence scores are obtained using a BERT-based \cite{devlin2019bert} sentiment analysis model.

\subsection{Time series forecasting model}

We build the forecaster architecture following Time-MMD \cite{liu2024time}. The primary distinction lies in our use of a reward-based fine-tuned LLM for summary generation, rather than directly incorporating the textual counterpart into the forecasting pipeline. As illustrated in Fig \ref{modelarchi}, the MMTSF pipeline consists of the following components: 1. A transformer architecture, based on \cite{zhou2021informer}, is trained on the numerical time series data to capture temporal patterns. 2. Textual data generated by LLM given prompt in Fig \ref{fig::multimodal_prompt} is processed using an open-source embedding model. We use the XLM-Roberta model \cite{conneau2019unsupervised} due to its strong multilingual capabilities. This model encodes the text into fixed-length embeddings. 3. A trainable projection layer maps the frozen embeddings into a space compatible with the numerical features. This layer amplifies the influence of the textual component while keeping the embedding model frozen to reduce computational overhead. The outputs of the projection layer $(\mathcal{Y}_{mlp})$ and the TSF model ($\mathcal{Y}_{tsf}$) are linearly combined with importance weight $\alpha$ to generate the final prediction: $\mathcal{Y} = \mathcal{Y}_{tsf} + \alpha \mathcal{Y}_{mlp}$.

\section{Emperical Evaluations}

\begin{figure}[h]
  \centering
  \includegraphics[width=\linewidth]{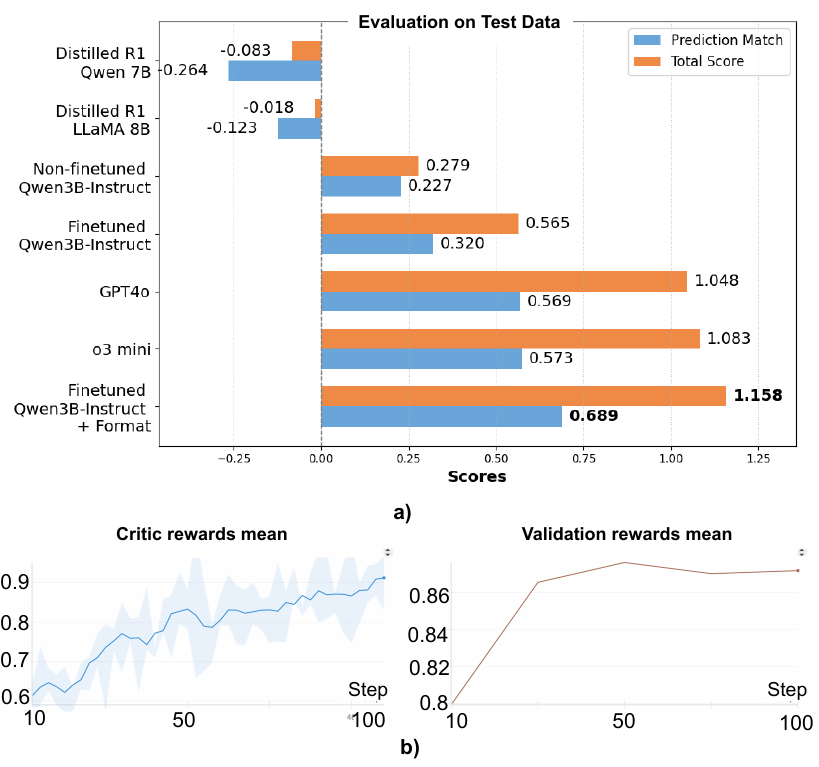}
  \caption{a) Evaluation scores on the test dataset b) Training c) Validation rewards during RL fine-tuning.}
  \label{rlresults}
\end{figure}

In this section, we present results from both the RL fine-tuning phase and the TSF pipeline. We select \textbf{Qwen2.5-3B Instruct} \cite{qwen2.5} as the LLM backbone due to its superior performance in the non-fine-tuned setting (see Fig.~\ref{rlresults}a) and low computational requirements for fine-tuning. We adopt the GRPO implementation from \cite{tinyzero}, which is built on top of the HybridFlow library \cite{sheng2024hybridflow}. The training reward and validation performance during RL fine-tuning are shown in Fig.~\ref{rlresults}b. The training time was 15.53 hours on two A100 GPUs costing approx USD60. Despite improvements through RL fine-tuning, the Qwen model does not always adhere to the expected output format. To address this, we apply a post-processing step using GPT-4o, which reformats Qwen's responses into the required structure. Importantly, GPT-4o only sees the raw response from Qwen—not the original query ensuring the integrity of the evaluation. This formatting of the output text can also be performed with any instruction following open source LLM.

We evaluate model performance on \emph{Prediction match} = format compliance + prediction accuracy, and the overall \emph{reward score} on the test dataset. As shown in Fig.~\ref{rlresults}a, the fine-tuned Qwen model with GPT-4o formatting achieves a \textbf{18.38\%} improvement in prediction accuracy and a \textbf{6.69\%} increase in the overall reward score compared to closed models such as o3-mini. Some models score negatively due to frequent formatting violations. A qualitative example of the output is presented in Table~\ref{tab:qualitative_example_full}. The non-finetuned Qwen simply repeated the prompt as output for this example. In Table ~\ref{tab:qualitative_example_full}, while both o3-mini and GPT-4o emphasize short-term recovery trends and predict an increase in clicks, they fail to account for strong early signal degradation due to significant keyword removals and CPA changes. In contrast, our model correctly interprets the impact of critical campaign changes, particularly the large-scale keyword removals, and predicts a decrease, aligning with the ground truth.

\begin{figure}[h]
  \centering
  \includegraphics[width=0.9\linewidth]{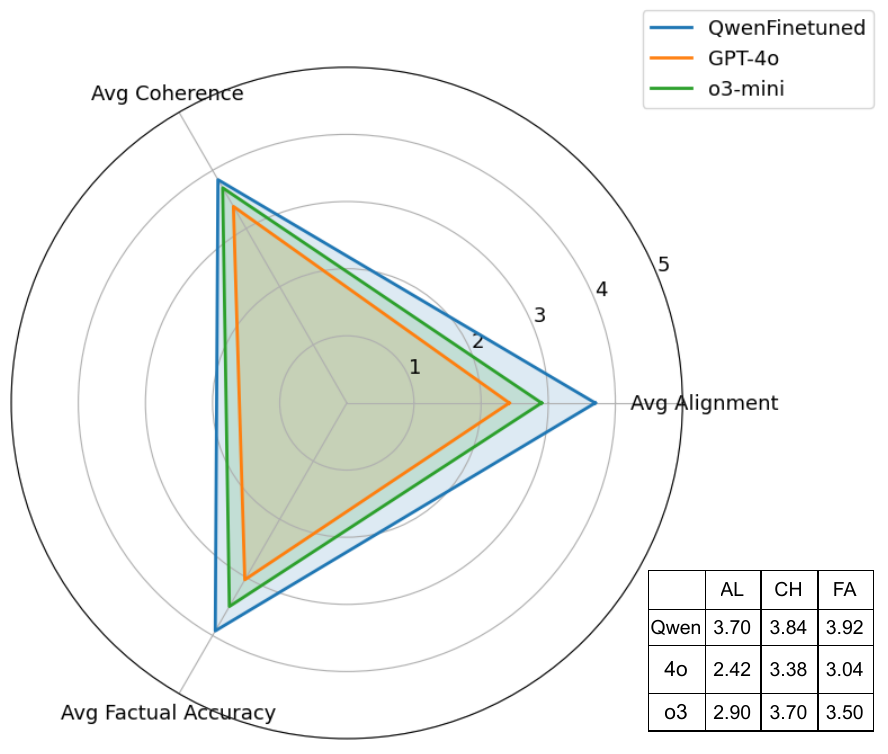}
  \caption{Scores for human evaluation w.r.t to Alignment (AL), Coherence (CH) and Factual Accuracy (FA).}
  \label{radarchart}
\end{figure}

\emph{We also conduct a human evaluation with five domain experts.} Each evaluator is presented with ten query highlights and corresponding model-generated reasonings from three models. They are asked to rate each reasoning on a scale of 1 to 5 (1 being the lowest) across three criteria:
1) \textbf{Alignment}: How well the reasoning aligns with the actual ground truth.
2) \textbf{Factual Accuracy}: How accurately the reasoning reflects the facts provided in the query.
3) \textbf{Coherence}: How logically consistent, well-structured and easy to follow the explanation is. The results of this evaluation are shown in Fig.~\ref{radarchart} \emph{with Qwen outperforming the other models. Fig.~\ref{radarchart} suggests that RL based finetuning not only produces more logically structured and readable explanations but also does a better job at grounding its reasoning and aligning it with actual outcomes.}

\begin{table}[htbp]
    \centering
    \caption{Qualitative Example of the reasoning and prediction provided by closed models and finetuned Qwen.}
    \small
    \begin{tabularx}{\linewidth}{@{}p{1.4cm}X@{}}
        \toprule
        \textbf{Query Highlights} & \textbf{Rolling Clicks (14 days)}: Day (0-6) declining from 0.787 to 0.444, Day (7-13) slight recovery to 0.509. \newline
        \textbf{Change logs}: Keyword Paused (Day 3), Keyword removals (Days 4,6,7), CPA increase (Day 5), Ad changes (Day 9,10). \newline
        \textbf{Ad Type}: SEARCH; \textbf{Bid Strategy}: MAXIMIZE\_CONVERSIONS.
        \newline
        \textbf{Ground Truth} : \textbf{Decrease}\\
        \midrule
        \textbf{Model} & \textbf{Reasoning \& Prediction}\\
        \midrule
        O3 Mini & <Reasoning>The rolling average of clicks exhibited a recovery after a downward trend, likely supported by the recent optimization changes such as ad text updates and asset creation. The adjustments seem to be stabilizing performance after early declines. </Reasoning><Prediction> Increase </Prediction>\\[3pt]
        GPT-4o & <Reasoning> The significant drop in clicks from Day 0 to Day 6 coincides with the removal of numerous keywords and an increase in target CPA, but the trend shows a slight recovery from Day 7 onwards. The recent changes in ad text and campaign assets may contribute to further improvement. </Reasoning><Prediction> Increase </Prediction>\\[3pt]
        \textbf{Qwen (Ours)} & <Reasoning> The rolling average of clicks shows a general downward trend from Day 0 to Day 6, followed by a slight recovery from Day 7 to Day 13. The significant removal of 113 phrase match keywords on Day 4 likely contributed to the initial decline in clicks. </Reasoning><Prediction> Decrease </Prediction>\\
        \bottomrule
    \end{tabularx}
    \label{tab:qualitative_example_full}
\end{table}

The TSF backbone is a Transformer Encoder model \cite{zhou2021informer}, configured with a 3-layer encoder, 4 attention heads, and a hidden size of 64. This is followed by a linear layer that outputs an $h$-step forecast. The amplification MLP operates on max pooled BERT-style text embeddings (dimension 768). It comprises three fully connected layers with ReLU activations and hidden sizes of [512, 256, 128], and it outputs an $h$-step amplification signal. For all experiments, we set the lookback window $l = 14$, the forecasting horizon $h = 5$, and the fusion weight $\alpha = 0.5$. We report the results of the full forecasting pipeline in Table~\ref{tab:comparison_methods}, using the following metrics: Mean Absolute Error (MAE) and Root Mean Squared Error (RMSE). The following baseline methods are compared:
1. \textbf{Copy:} Predicts the next $h$ steps by repeating the last $h$ days average in the look-back window. 2. \textbf{Uni:} Uses only the trained numeric forecaster. 3. \textbf{Multi + change-log:} Incorporates raw change log text. 4. \textbf{Multi + GPT-4o:} Uses summaries generated by GPT-4o as the textual component.
We omit experiments with o3-mini-based summaries due to high API costs and the marginal difference in performance compared to GPT-4o, as shown in Fig.~\ref{rlresults}a. As shown in Table~\ref{tab:comparison_methods}, our multimodal forecaster achieves a lower MAE and RMSE than all baselines, demonstrating the effectiveness of our model.

\begin{table}[htbp]
    \centering
    \small
    \caption{Performance Comparison of Forecasting Methods (mean ± std) over 3 random seeds scaled by $10^{2}$.}
    \begin{tabularx}{\linewidth}{lXXX}
        \toprule
        \textbf{Method} & \textbf{MAE ($\downarrow$)} & \textbf{RMSE ($\downarrow$)} \\
        \midrule
        Copy & 7.031  & 10.373 \\
        Uni & $5.102 \pm 0.07$ &  $7.800 \pm 0.002$ \\
        Multi with Changelog & $5.379 \pm 0.01$ &  $8.051 \pm 0.001$ \\
        Multi with GPT-4o & $5.034 \pm 0.03$ &  $7.700 \pm 0.001$ \\
        \textbf{Multi with Qwen (Ours)} & $\mathbf{4.948}$ $\pm$ 0.02 & $\mathbf{7.670}$ $\pm$ 0.001 \\
        \bottomrule
    \end{tabularx}
    \label{tab:comparison_methods}
\end{table}

\section{Conclusion and Future Work}

In this paper, we investigate multimodal click prediction for digital advertising. We propose a novel reward function to fine-tune LLM using real-world advertisement data. Furthermore, we present an end-to-end MMTSF pipeline that integrates both textual summarization and numerical time series data, demonstrating improved predictive performance. To the best of our knowledge, this is the first work to incorporate textual reasoning into time series forecasting. There remains significant potential for enhancing the reward design to make the generated reasoning more logically consistent and aligned with forecasting outcomes. In future work, we aim to explore this direction and develop more structured reasoning mechanisms within multimodal forecasting frameworks.

\section{GenAI Usage Disclosure}

ChatGpt-4o was used to polish the writing of this manuscript. Github copilot was used as a coding assistant tool.

%% \begin{acks}
%% To Robert, for the bagels and explaining CMYK and color spaces.
%% \end{acks}

\bibliographystyle{ACM-Reference-Format}
\balance
\bibliography{sample-base}

\appendix

\end{document}